\documentclass[conference]{IEEEtran}
\usepackage{amsmath, soul}
\usepackage{subcaption}
\usepackage{booktabs}
\usepackage{flushend}
\usepackage{float}
\usepackage{graphicx}
\usepackage{multirow}
\usepackage{makecell}

\newcolumntype{P}[1]{>{\centering\arraybackslash}p{#1}}

\begin{document}
\title{The Impact of Projection and Backboning\\on Network Topologies}

\author{\IEEEauthorblockN{Michele Coscia}
\IEEEauthorblockA{\textit{IT University of Copenhagen}}
\textit{Copenhagen, DK} \\
mcos@itu.dk
\and
\IEEEauthorblockN{Luca Rossi}
\IEEEauthorblockA{\textit{IT University of Copenhagen}}
\textit{Copenhagen, DK}\\
lucr@itu.dk
}

\maketitle

\begin{abstract}
Bipartite networks are a well known strategy to study a variety of phenomena. The commonly used method to deal with this type of network is to project the bipartite data into a unipartite weighted graph and then using a backboning technique to extract only the meaningful edges. Despite the wide availability of different methods both for projection and backboning, we believe that there has been little attention to the effect that the combination of these two processes has on the data and on the resulting network topology. In this paper we study the effect that the possible combinations of projection and backboning techniques have on a bipartite network. We show that the 12 methods group into two clusters producing unipartite networks with very different topologies. We also show that the resulting level of network centralization is highly affected by the combination of projection and backboning applied.
\end{abstract}

\section{Introduction}
Over the years bipartite networks, network composed of two mutually exclusive set of nodes where edges can exist only between nodes belonging to different sets, have emerged as a fruitful way to study a number of phenomena \cite{neal2014backbone}. Also known as affiliation networks or two-mode networks \cite{wasserman1994social}, bipartite networks have been used to study participation in social events \cite{davis2009deep}, individuals sitting on multiple corporate boards \cite{mizruchi1996interlocks}, the movie industry \cite{watts1998collective}, academic productivity \cite{ductor2015does}, the legislative process \cite{fowler2006legislative} and the evolution of tastes and preferences\cite{lambiotte2005uncovering}.

While methods are emerging to study bipartite networks as such \cite{agneessens2013introduction, lehmann2008biclique}, it is fair to say that the majority of research developed with bipartite data is done through a preliminary projection of the bipartite network and then the application of a backboning technique on the resulting weighted network to extract only the most significant edges \cite{neal2014backbone}. Despite this can be considered a common research practice, we believe that there has been little research, so far, to understand the consequences of the interplay between projection methods and backboning techniques. To what extent, and in which direction, will a specific projection method together with a specific backboning process influence the structural properties of the resulting unipartite network?

This paper is, to the best of out knowledge, the first attempt to address this issue systematically. We do not aim at providing a comprehensive review of all the existing projection and backboning methods (see \cite{pavlopoulos2018bipartite} for a review of the available methods in the life-sciences), even if we have tried to be conceptually exhaustive, but rather to provide evidences of the joint effect of the methods on the resulting networks.

We define a strategy as a specific combination of projection and backboning methods. We find that, despite the large number of possible strategies that we have tested (12), the resulting networks cluster into two groups according to their topological properties. We also show how the main difference between those clusters is the level of centralization of the resulting network, with a group of strategies producing decentralized networks and the other group producing centralized topologies. We conclude by stressing the importance of this result and calling for more transparency about the effects that the research strategy we adopt might have on the results we report.

\section{Related Work}\label{sec:related}
Despite its apparent simplicity, there are two aspects, in this way of dealing with bipartite networks, that should be stressed in order to fully understand the process behind the study of projected unipartite networks and its underlying assumptions.

First, bipartite networks are used to represent a wide range of diverse scenarios and projecting those bipartite networks build on different, often very different, assumptions. As Wassermann and Faust \cite{wasserman1994social} have extensively discussed we should acknowledge a difference between affiliation (membership) networks and collaboration networks. In affiliation networks the observed participation of two actors to the same social event assumes the existence of a connection between those actors (or it assumes that the probability of that connection to exist increases). In collaboration networks we use the actual output of an existing social connection (e.g. a scientific publication) to infer the relation between the actors. Yet another case is represented by taste-network where the preference expressed by an actor toward a specific object (e.g. a music genre) is projected into a connection expressing similarity between the actor and another actor with similar tastes \cite{zhou2007bipartite}. All these different social processes that exist underlying the original bipartite data are removed by the projection into a unipartite network.

Second, in the process of projecting a bipartite network, information is always lost. Usually quantitative information about frequency -- e.g. how many events two actors have co-attended -- is preserved as a weight of the edge. We will discuss various strategies to do so, in the following sections. However, qualitative information about how nodes were connected before the projection -- e.g. the actual events the actors co-attended -- is usually lost, suggesting caution during the interpretation of the resulting networks.

Among the many areas that have used bipartite data, the present paper was inspired by a recent study published in the area of communication research, more precisely in the sub-field of online audience studies \cite{majo2018backbone}. In this context, audience networks have been used for several years \cite{ksiazek2011network} to study the general structure of audiences and answer key questions about echo-chambers, the existence of trend toward cyber-balkanization and filter bubbles. Audience networks are networks where nodes represents media sites that are connected if they share at least a part of their audience \cite{majo2018backbone}. While there are several strategies to collect audience behaviour, over the recent years the common practice has been the use of trace data -- digital traces left behind by users browsing the web \cite{majo2018backbone}. This is also due to intrinsic limitation of self-reported data \cite{prior2009immensely}.

Within this perspective, audience networks are the projection of a bipartite network connecting each user with the media-sites she consumes. In the existing literature there seem to exist two different approaches to realize these projected networks. (I) Once the existence of a shared audience between two sites has been observed, only the shared audience that is greater than an expected level of random chance is projected as an edge between the two media sites \cite{taneja2015global}. Alternatively, (II) the user-media source bipartite network is simply projected with the edge weights representing the level of audience overlap between those sites, then backboning is performed to extract only the statistically significant edges \cite{majo2018backbone}. The two approaches deal with a similar set of problems, but they do so in different ways. While at the time of writing there is an ongoing debate between the authors supporting different approaches \cite{webster2018building}, we claim that in both cases little attention has been given to the consequences of the projection and backboning process, and to two specific problems: which projection and backboning method should be used given the data we are dealing with, and what is the effect of the chosen procedure on the resulting network.

\section{Problem Definition}\label{sec:problem}

To observe the impact of different projection and backboning methods on the network topology we first define a strategy to compare the resulting networks, then, in Section \ref{sec:algorithms} we apply a combination of three different projection methods and four backboning methods to a real-word network conceptually similar to what has been used in the article that inspired us (described in Section \ref{sec:exp-data} and then, in Section \ref{sec:impact} we discuss the results). Given our focus on the network topology we will measure the similarity between the networks resulting from the projection-backboning strategies through four different methods: neighbor Jaccard, clustering coefficient, degree correlation and centralization.

\subsection{Estimating Network Similarity}

\subsubsection{Neighbor Jaccard} 
We first take into consideration the neighbors of the nodes in the resulting networks. We do so by using Jaccard coefficient. Jaccard is a well known measure and we modify it as follows:

\[J(u,G_1,G_2)=\frac{|\Gamma_{G_1}(u) \cap \Gamma_{G_2}(u)|}{|\Gamma_{G_1}(u) \cup \Gamma_{G_2}(u)|}\]

where \(\Gamma_{G_1}(u)\) denotes the set of neighbors of \(u\) in graph $G_1$. In practice, this formula tells us how similar the neighborhood of node $u$ is in the two graphs $G_1$ and $G_2$ -- in our case extracted using projection/backboning techniques. By averaging $J(u,G_1,G_2)$ across all $u$s in our network, we can estimate how similar $G_1$ and $G_2$ are:

$$ sim(G_1, G_2) = \dfrac{1}{|V|} \sum \limits_{u \in V} J(u,G_1,G_2),$$

with $V$ being the (shared) set of nodes in $G_1$ and $G_2$.

\subsubsection{Clustering Coefficient}
One of the fundamental characteristics of real world networks is that they are clustered. This means that the number of closed triangles is larger than null expectation. In terms of social network, this means that ``the friend of my friend is likely to be my friend''. The clustering coefficient captures this tendency \cite{newman2003structure}:

$$ CC = \dfrac{3 \times \#Triangles}{\#Triads}.$$

The clustering coefficient is one of the most salient simple statistics describing the structure of a network. We transform this measure in a similarity measure by computing the absolute value of the difference of the coefficients for two networks.

\subsubsection{Degree Correlation}
Let us assume that we performed two alternative bipartite projections and backboning and obtained graphs $G_1$ and $G_2$. The same node $u$ in $G_1$ and $G_2$ will have two different degrees: $|\Gamma_{G_1}(u)|$ and $|\Gamma_{G_2}(u)|$. In this method, we create a degree vector per network and we calculate the Spearman correlation of these vectors.

This test ensures that the hubs (peripheral) nodes in one projection are also hubs (peripheral) in the other. We use the Spearman correlation, because it is not sensitive to the typical skewedness of degree distributions \cite{barabasi2003scale}.

\subsubsection{Centralization}
As the last measure to compare the resulting graphs we use centralization based on the Betweenness Centrality. Centralization is a method originally introduced by Freeman \cite{freeman1978centrality} to measure how centralized is the whole network, and it is defined as: 

\[\sum_{v}(max_w c_w - c_v),\]

where $c_v$ is the betweenness centrality of node $v$. In practice, centralization compares the difference between the most central node in the network with the one of all other nodes in the network. This is then compared to the maximally centralized possible network with the same number of nodes as the original network, usually a star.

\section{Algorithms}\label{sec:algorithms}

\subsection{Network Projection}\label{sec:algorithms-projection}

\begin{figure}
\centering
\includegraphics[width=.66\columnwidth]{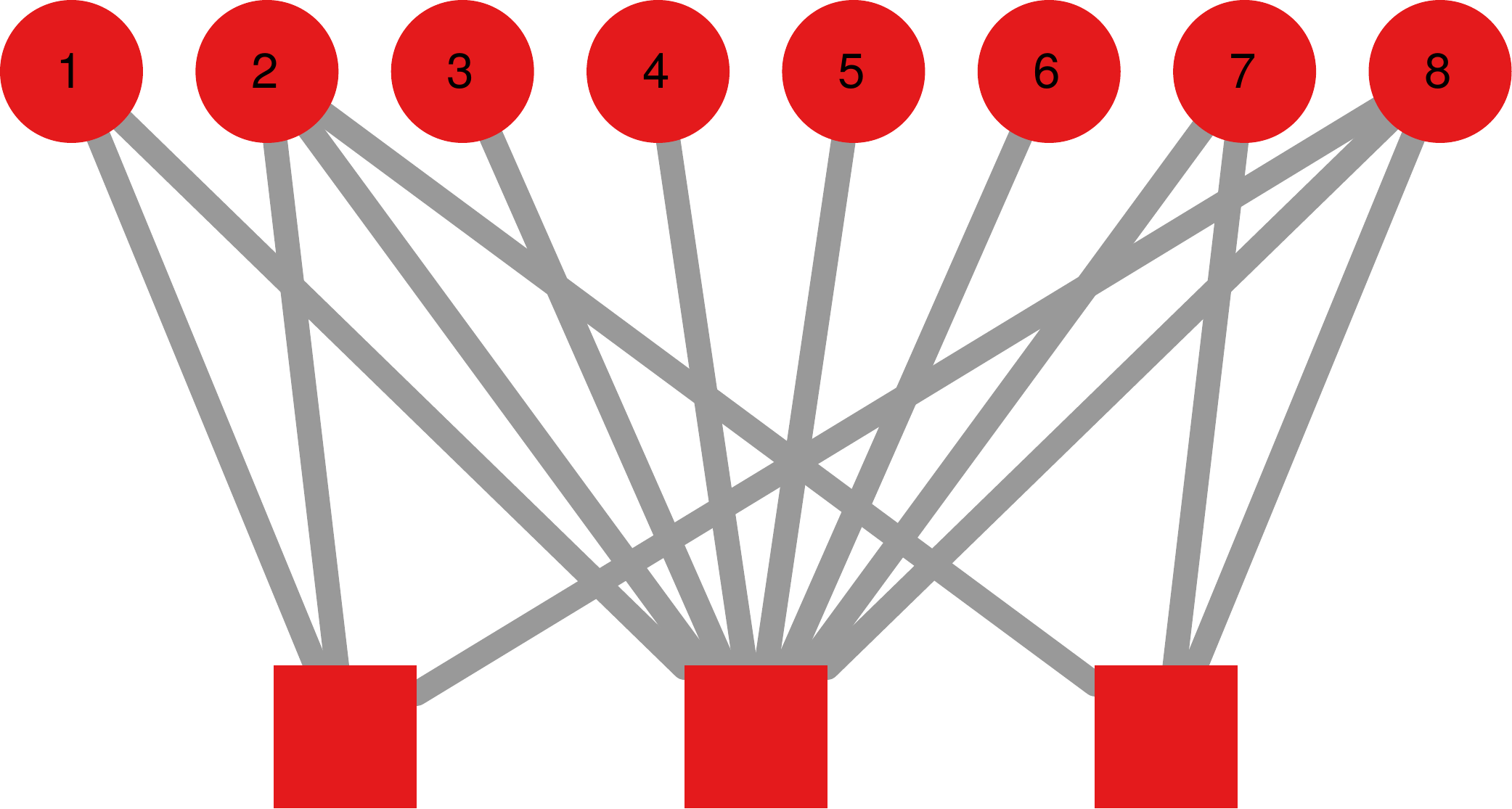}
\caption{The toy bipartite network for our examples.}
\label{fig:projections}
\end{figure}

The aim of a network projection technique is to take a bipartite network and find reasonable edge weights for its unipartite projection. Assume that the bipartite network has two node types, $V_1$ and $V_2$, thus $G = (V_1, V_2, E)$. A unipartite projection only includes nodes in either of those classes, say $V_1$. The original edges between $V_1$ and $V_2$ nodes are used to establish the weights in the projection. Figure \ref{fig:projections} represents the toy bipartite network we use for the examples in this section.

\begin{figure}
\centering
\includegraphics[width=.36\columnwidth]{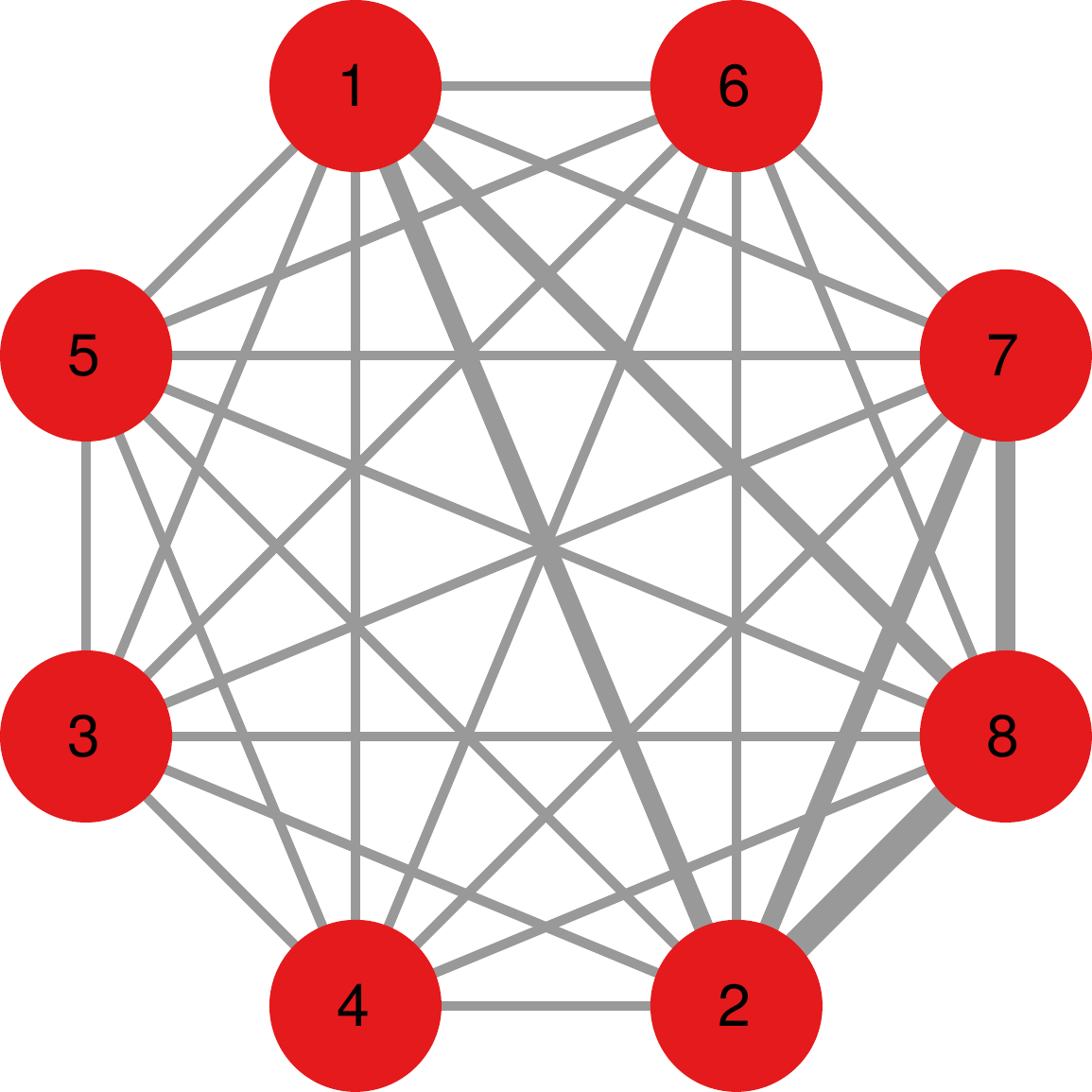}\qquad\qquad
\includegraphics[width=.36\columnwidth]{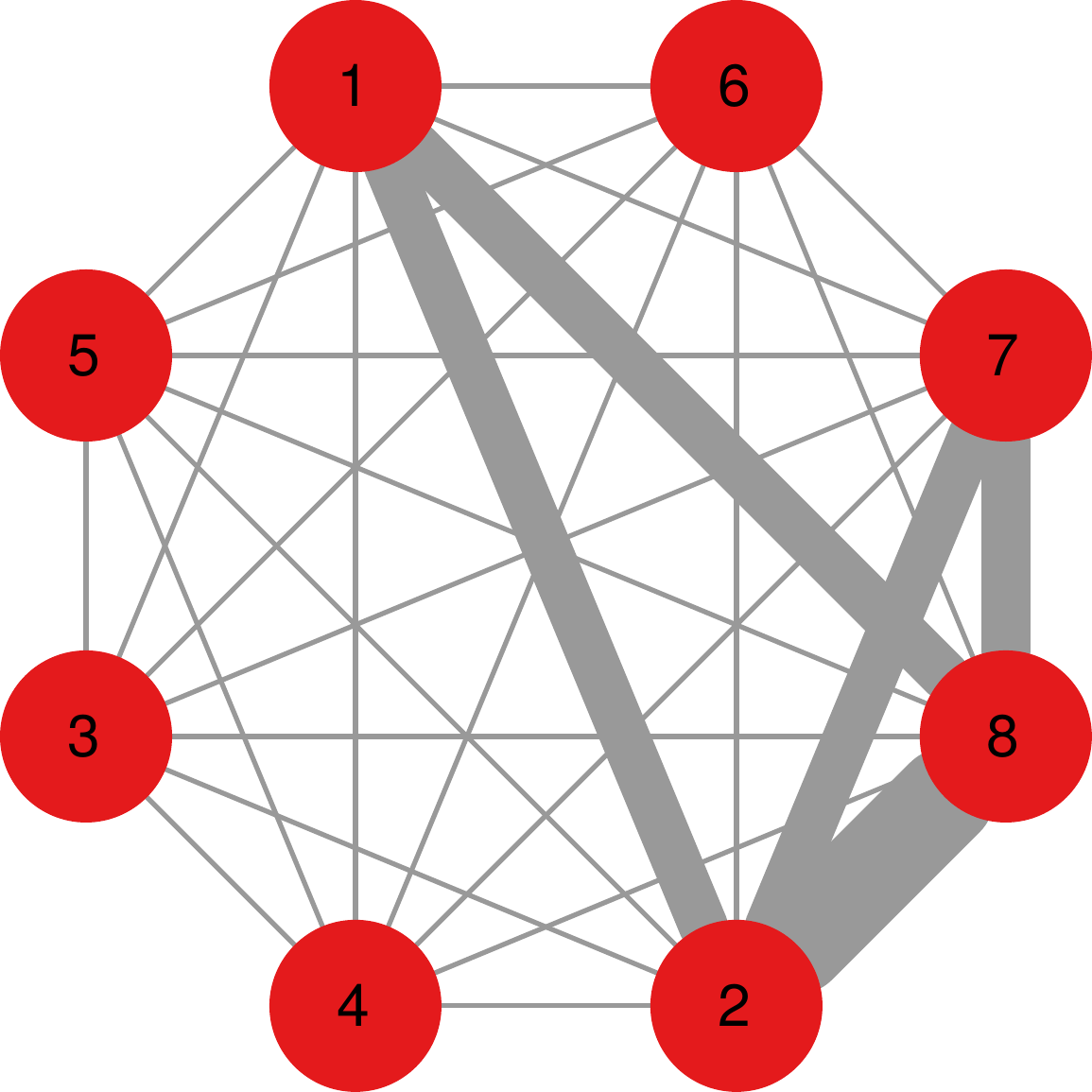}
\caption{The results of two bipartite projection techniques. Edge widths are proportional to the resulting weights. (Left) Simple. (Right) Hyperbolic.}
\label{fig:projections1}
\end{figure}

The simplest approach to network projection is to simply count the number of common neighbors between two nodes in $V_1$. Say that $\Gamma(u)$ represents the set of neighbors of node $u$ in the original bipartite structure. The simple projection says that the weight between nodes $u,v \in V_1$ is $w_{u,v} =  |\Gamma(u) \cap \Gamma(v)$. Figure \ref{fig:projections1}(left) shows the result of the simple projection to the bipartite network in Figure \ref{fig:projections}.

The evident downside of the simple projection is that, in most cases, the differences between edges are small and discrete. This makes the usual following step, network backboning, challenging. Bipartite networks with broad and correlated degree distributions will generate skewed unipartite edge weights distributions which are not easy to treat.

On the basis of this observation, we consider three popular alternatives for network projection: Hyperbolic \cite{newman2001scientific}, ProbS \cite{zhou2007bipartite}, YCN \cite{yildirim2014using}.

When developing the hyperbolic projection, Newman observed that the relationship between two sole co-authors of a paper is stronger than the one between two authors co-authoring a paper with a thousand more other scientists \cite{newman2001scientific}. More formally, the degree of the $V_2$ node should discount its contribution to $w_{u,v}$.

Newman named his projection ``hyperbolic'' because he chose to discount the contribution by the inverse of the degree: $w_{u,v} =  \sum \limits_{z \in \Gamma(u) \cap \Gamma(v)} |\Gamma(z)|^{-1}$. Figure \ref{fig:projections1}(right) shows the result of the hyperbolic projection to the bipartite network in Figure \ref{fig:projections}. As we can see, hyperbolic greatly exaggerates the difference in contribution between the square nodes with $|\Gamma(z)| = 3$ and the one with $|\Gamma(z)| = 8$.

\begin{figure}
\centering
\includegraphics[width=.36\columnwidth]{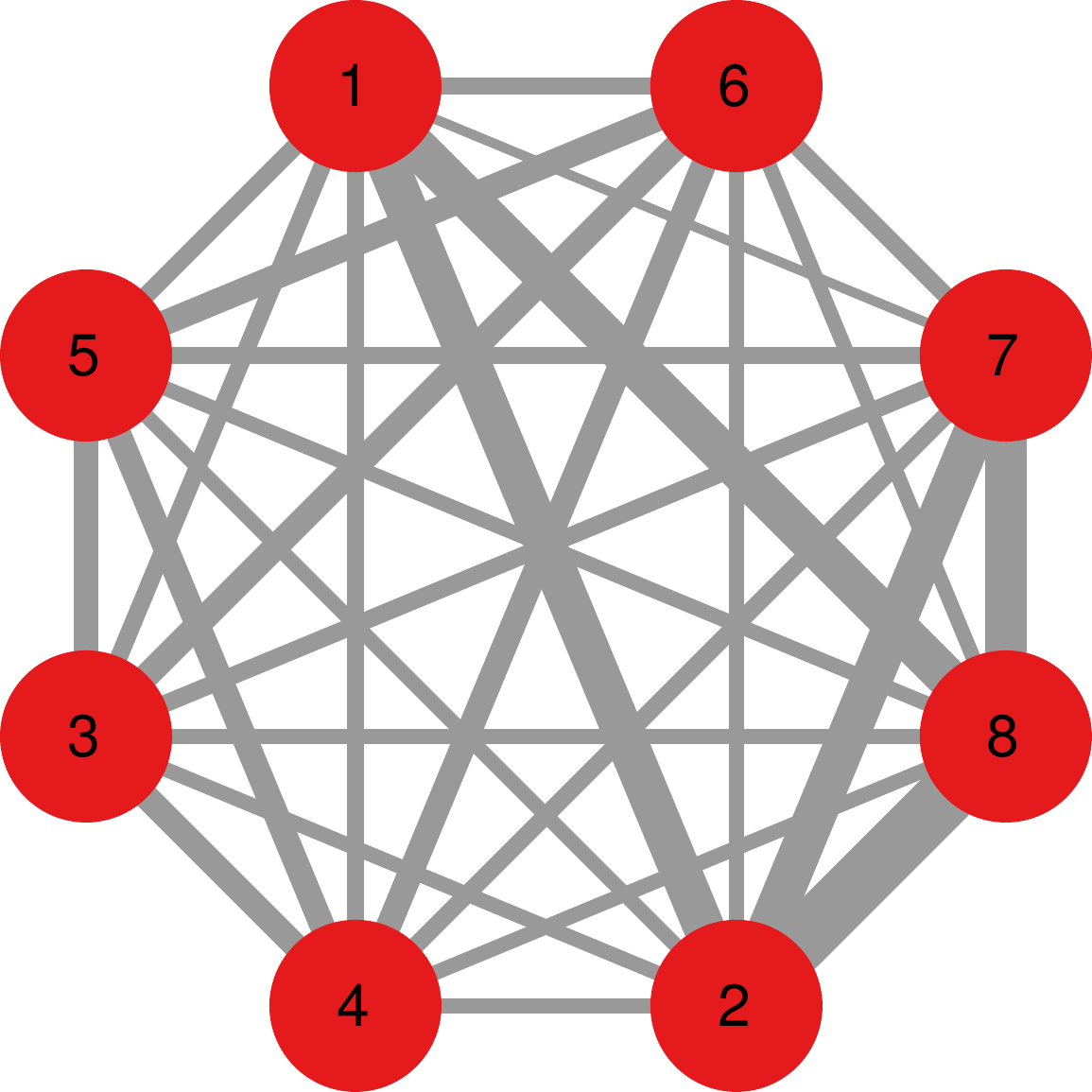}\qquad\qquad
\includegraphics[width=.36\columnwidth]{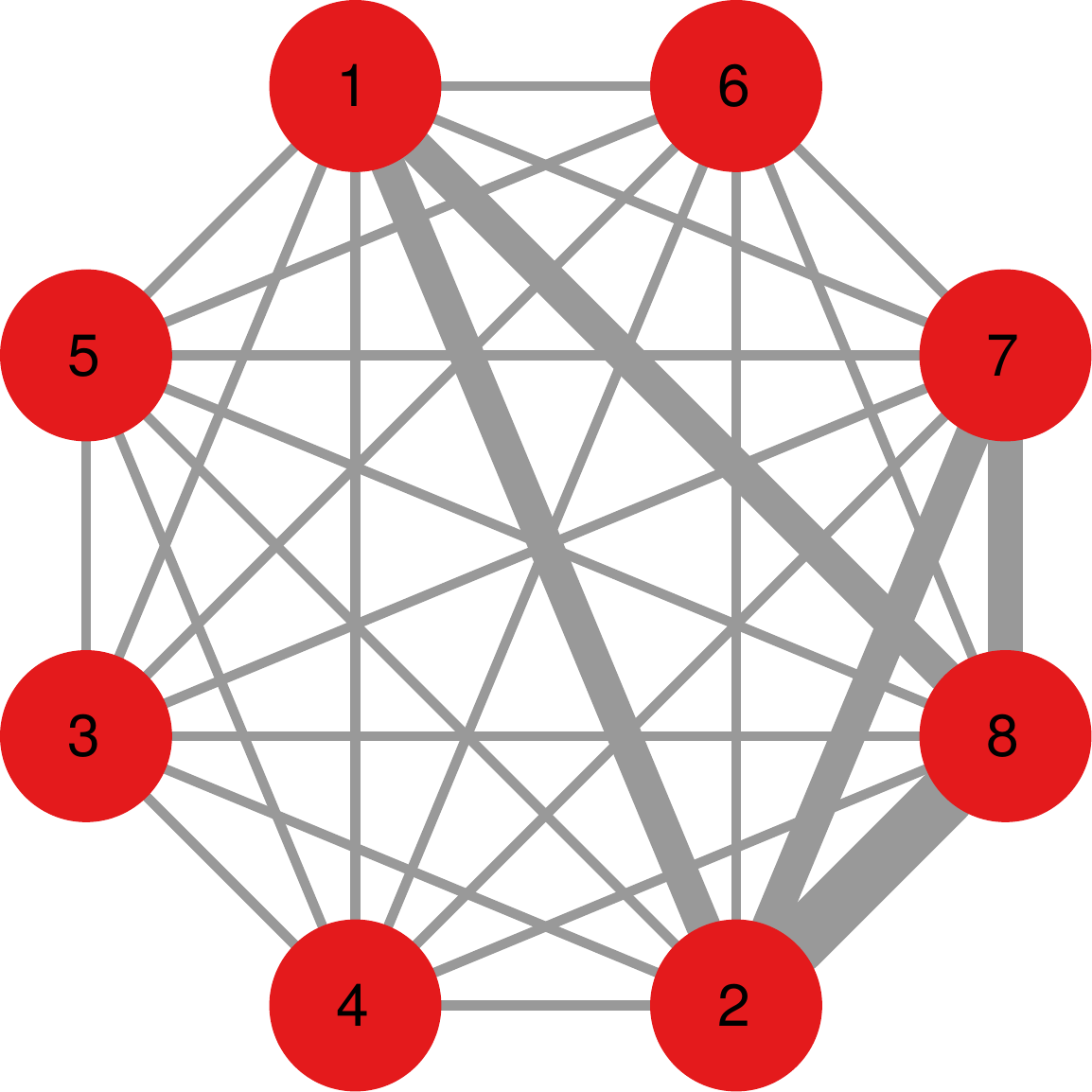}
\caption{The results of two bipartite projection techniques. Edge widths are proportional to the resulting weights. (Left) ProbS. (Right) YCN.}
\label{fig:projections2}
\end{figure}

ProbS \cite{zhou2007bipartite} -- or ``resource allocation'' -- applies the same reasoning in both directions. Not only a paper with a thousand authors contributes less to the relationship between two authors: authors with thousands of papers distribute their relationship strength equally among them. So the strength between $u$ sends to $v$ via $z$ depends on $\Gamma(u)$ and $\Gamma(z)$. Formally: $w_{u,v} =  \sum \limits_{z \in \Gamma(u) \cap \Gamma(v)} (|\Gamma(u)||\Gamma(z)|)^{-1}$.

Note that, in the vast majority of cases, $|\Gamma(u)| \neq |\Gamma(v)|$: the nodes we are trying to connect do not have the same degree. This implies that $w_{u,v} \neq w_{v,u}$. Thus ProbS generates directed unipartite projections. Since in this paper we are interested in symmetric relationships, we average the $w_{u,v}$ and $w_{v,u}$ values for the edge connecting $u$ and $v$, rendering ProbS symmetric. This is equivalent of using the Hybrid method with $\lambda = 0.5$, since that would be the average between ProbS and its transpose \cite{zhou2010solving}.

Figure \ref{fig:projections2}(left) shows the result of the ProbS projection to the bipartite network in Figure \ref{fig:projections}. Note the fundamental difference between ProbS and the previous approaches: $V_1$ nodes connect more strongly to each other than in the hyperbolic even if they share a $V_2$ node with high degree, due to their own low degree.

YCN is an evolution of ProbS \cite{yildirim2014using}. In ProbS, all that matters is the degree of the $V_1$ and $V_2$ nodes. One could consider ProbS as performing random walks of length two and asking the probability of ending in node $v$ when starting from node $u$. YCN, rather than stopping at length two, performs infinite length random walks. Thus, it uses the stationary distribution $\pi$ of $AA^T$, assuming $A$ is the adjacency matrix of $G$ ($AA^T$ is the basic structure used in ProbS, which uses no additional information).

Figure \ref{fig:projections2}(left) shows the result of the YCN projection to the bipartite network in Figure \ref{fig:projections}. It highlights how YCN tends to exaggerate the differences between nodes less than the hyperbolic projection.

\subsection{Network Backboning}\label{sec:algorithms-backboning}
The aim of network backboning is to take a dense weighted network whose edge weights are possibly affected by noisy measurements and return a sparser network devoid of noise, which is more tractable by standard network algorithms.

The most common approach to solve this problem is establishing a threshold. All edges whose weight is higher than the threshold are kept, while all other edges are dropped. This is not a great idea in case of degree correlations and skewed weight distributions. Both conditions are extremely common in real world networks \cite{serrano2009extracting}.

Degree correlations mean that a fixed threshold might remove all connections in some parts of the networks while removing none in another part. Skewed weight distributions make it hard to find a proper threshold, as low values still prune the majority of edges, and to motivate it, as power law distributions might have undefined variance.

To fix these problems several network backboning approaches have been proposed. We focus on two state of the art algorithms: Disparity Filter (DF) \cite{serrano2009extracting} and Noise-Corrected (NC) \cite{coscia2017network}.

\begin{figure}
\centering
\includegraphics[width=.45\columnwidth]{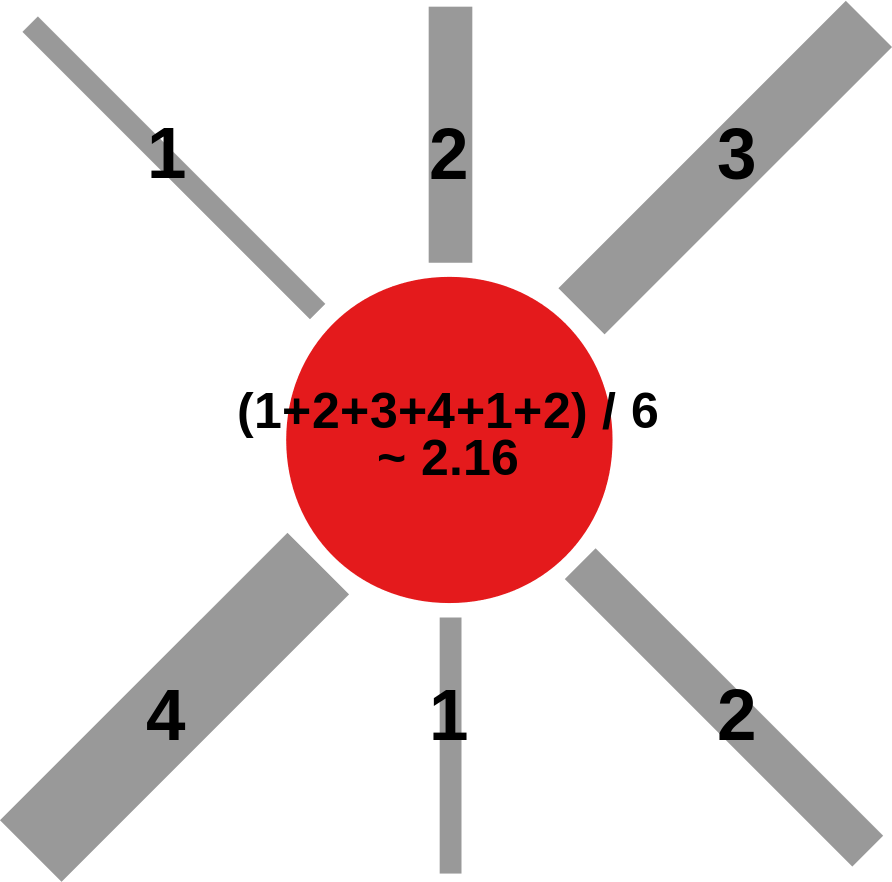}\qquad
\includegraphics[width=.45\columnwidth]{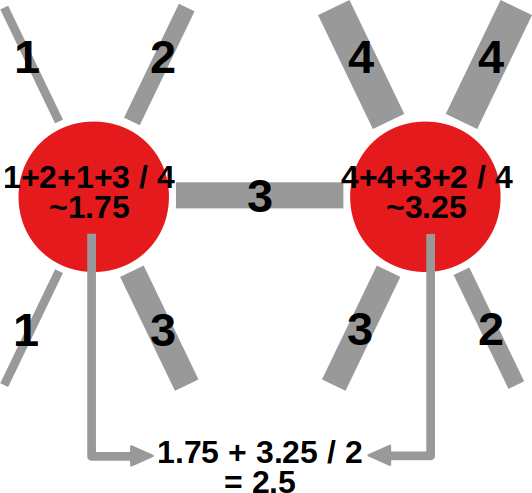}
\caption{(Left) A simplification of DF backboning. The node determines its expected edge weight and keeps only the edges significantly heavier than this expectation, in the example all edges with weight higher than 2.16. (Right) A simplification of NC backboning. The edge its expected weight and is kept only if it exceeds this expectation. In the example, the edge is kept because observation (3) is greater than expectation (2.5).}
\label{fig:backboning1}
\end{figure}

Both approaches re-weight the edges such that thresholds defined in this new space are meaningful. DF spreads weights over a uniform distribution null assumption. For each edge weight it asks whether the weight is significant given the expected (average) weight of all edges of either nodes. Figure \ref{fig:backboning1}(left) shows a simplification of the DF's approach.

Given that high degree nodes tend to connect more strongly than peripheral nodes, DF tends to create a hub-spoke structure, because the weight of an edge connecting to a hub will likely to be significant for a normally weak peripheral node. 

On the other hand, NC sees the edge as a collaboration between the two connected nodes. As a result, it creates not a node-dependent expectation, but an edge-dependent one. It asks whether the edge weight is significant when considering an interaction between both nodes'  average edge weights. Figure \ref{fig:backboning1}(right) shows a simplification of the NC's approach.

\begin{figure}
\centering
\includegraphics[width=.66\columnwidth]{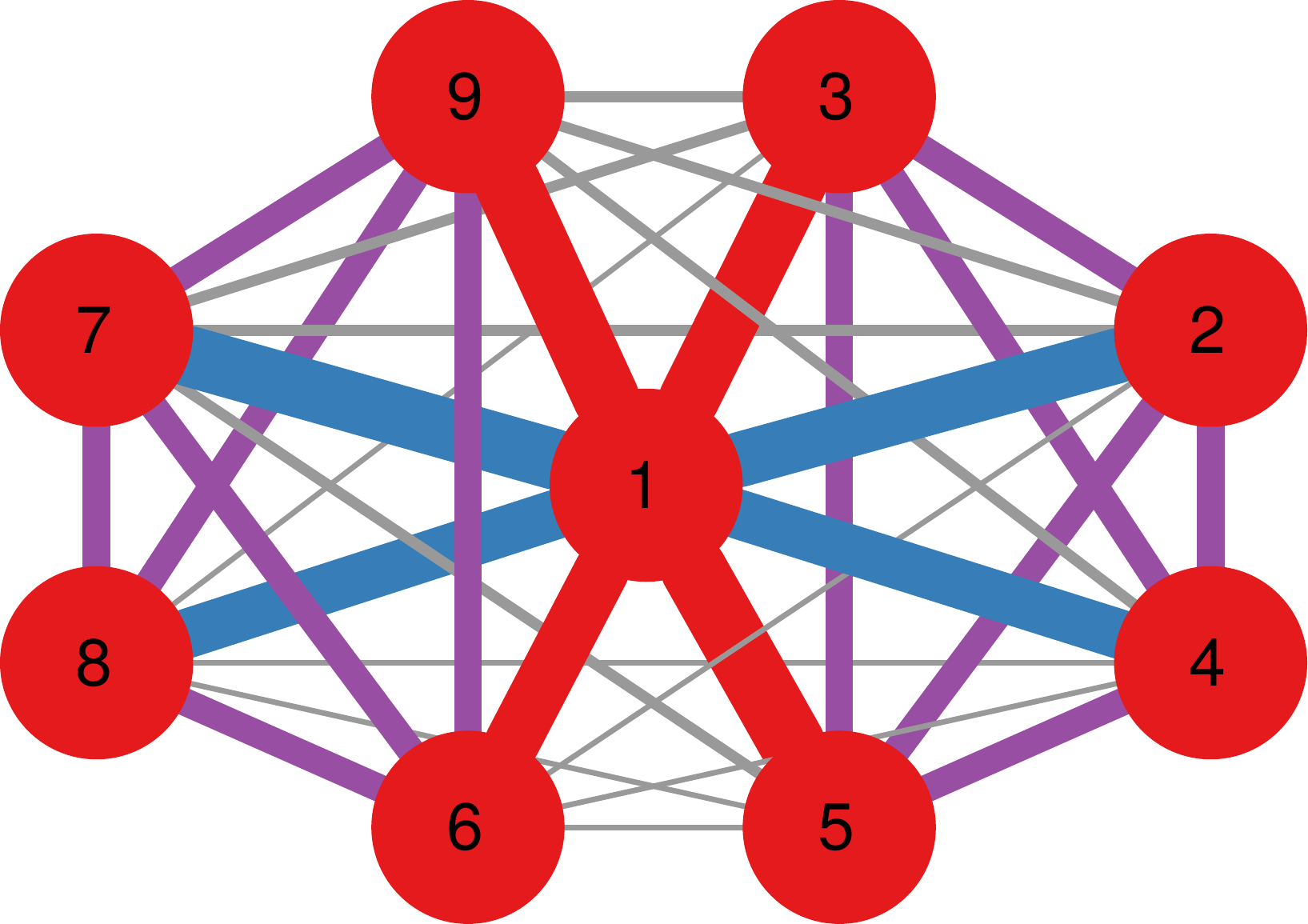}
\caption{Different choices between DF and NC backboning. Edge width is proportional to its weight. Edge color: red = selected by both DF and NC; blue = DF only; purple = NC only; gray = neither.}
\label{fig:backboning2}
\end{figure}

As a result, NC tends to return a decentralized community structure, rather than a strong core-periphery topology as DF. Figure \ref{fig:backboning2} shows a comparison between the two algorithms, highlighting their different choices. DF selects only the edges of node 1, because, as a hub, it generally connects more strongly with everything. NC recognizes node 1's connections only when they are stronger than what node 1 usually sends, and so it prefers to keep the connections in the (2,3,4,5) and (6,7,8,9) communities, which are stronger than what those nodes generally send outside their community.

\section{Case Study}\label{sec:exp}
In this section, we apply the projection-backboning framework to generate a unipartite network from bipartite data. While our work was originally inspired by \cite{majo2018backbone} we did not have access to ComScore data. Instead we reconstructed an audience network using Twitter data collected by  the Mapping Italian News Media Political Coverage in the Lead-up of 2018 General Election \cite{giglietto2018mapping}. The project released the tweets (ids) produced during February and March 2018 by the top 5000 contributors of political retweets on the website \texttt{elezioni2018.news}.

\subsection{Raw Data}\label{sec:exp-data}

\begin{table}
    \centering
    \begin{tabular}{l|r}
    Statistic & Value  \\
    \hline
    \# Tweets & 4,385,877\\
    \# Users & 3,853\\
    \# Domains & 19,108\\
    \# Edges & 175,329\\
    \end{tabular}
    \caption{The basic statistics of our dataset.}
    \label{tab:basicstats}
\end{table}

From these data we generate the user-domain bipartite network where the users are the authors of the tweets and the domains are the domains shared in the text. Table \ref{tab:basicstats} contains basic statistics about our dataset. Note that the count of users in the bipartite network is lower than 5,000, because some users never shared a domain different than \texttt{twitter.com}, which we remove as it is greatly overexpressed in the data.

\begin{figure}
\centering
\includegraphics[width=.49\columnwidth]{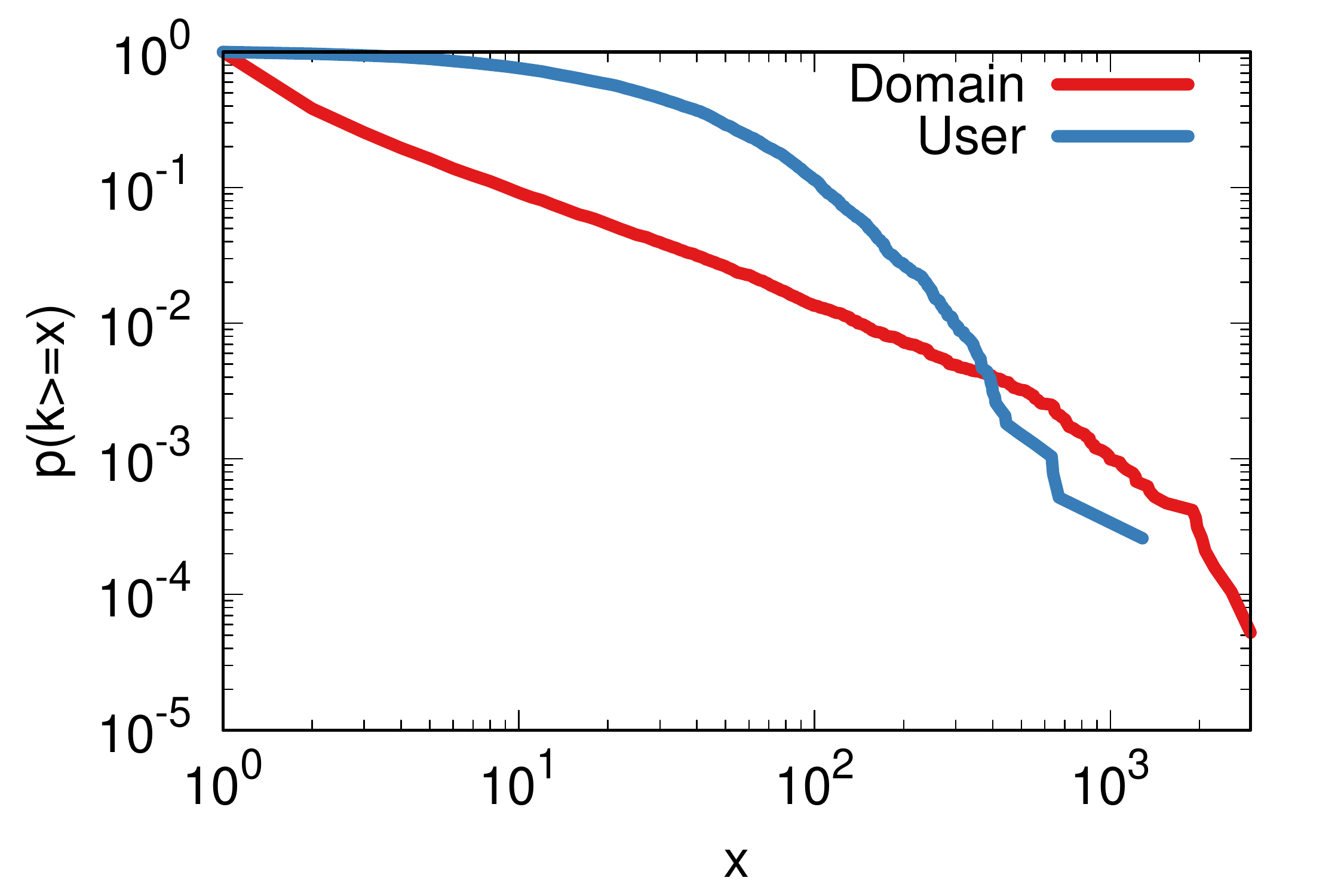}
\includegraphics[width=.49\columnwidth]{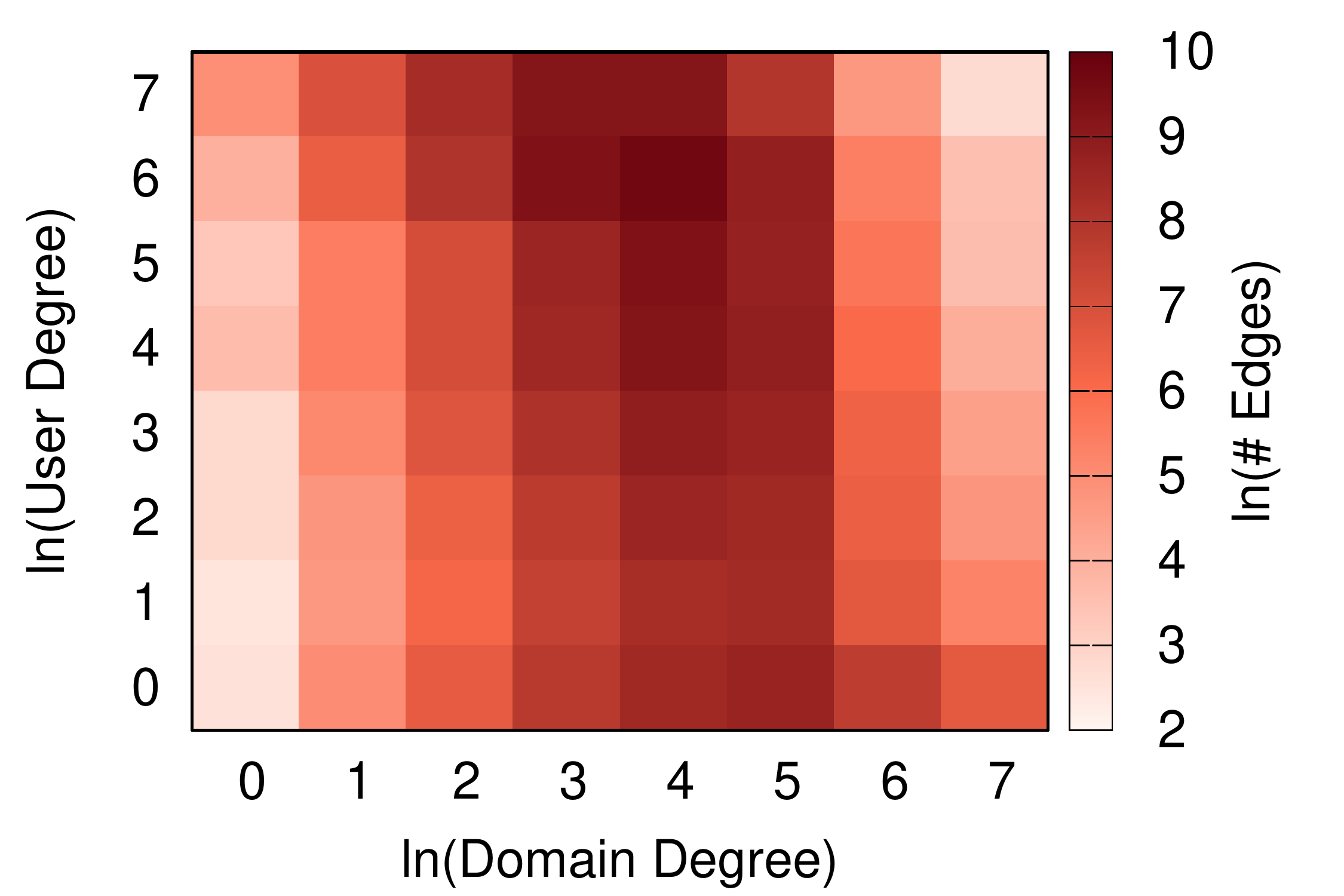}
\caption{(Left) The cumulative degree distribution of domains (red) and users (blue) in log-log scale. (Right) The joint domain-user degree distribution: how many edges (color) connects domains with a given degree (x-axis, log scale) to users of a given degree (y-axis, log scale).}
\label{fig:degdistr}
\end{figure}

\begin{figure*}
\centering
\includegraphics[width=.475\columnwidth]{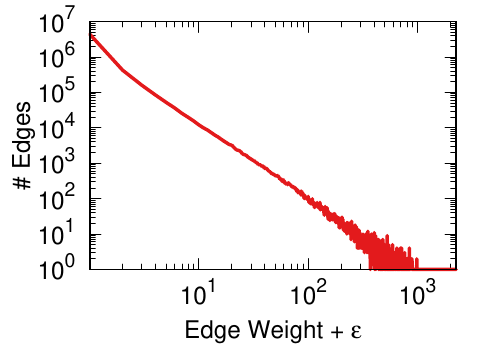}
\includegraphics[width=.475\columnwidth]{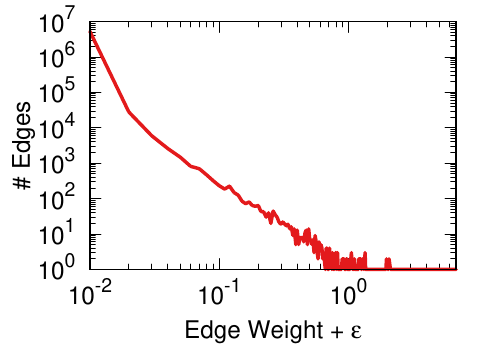}
\includegraphics[width=.475\columnwidth]{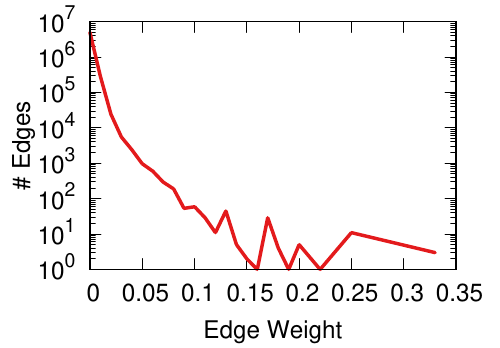}
\includegraphics[width=.475\columnwidth]{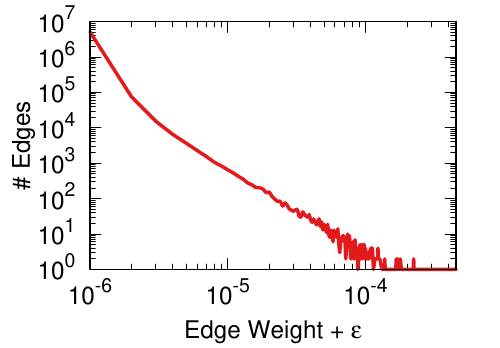}

\includegraphics[width=.475\columnwidth]{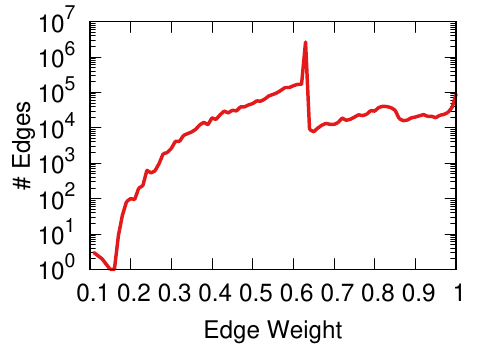}
\includegraphics[width=.475\columnwidth]{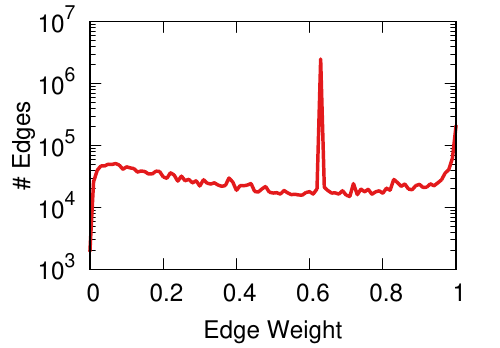}
\includegraphics[width=.475\columnwidth]{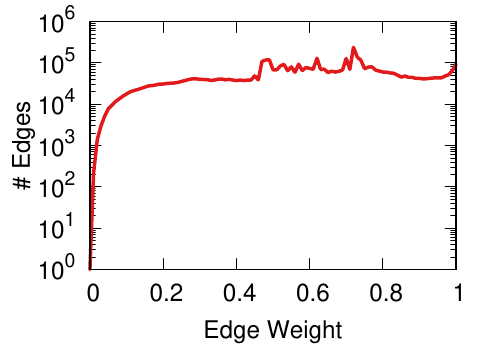}
\includegraphics[width=.475\columnwidth]{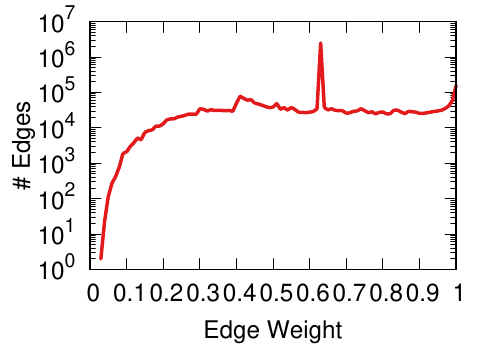}

\includegraphics[width=.475\columnwidth]{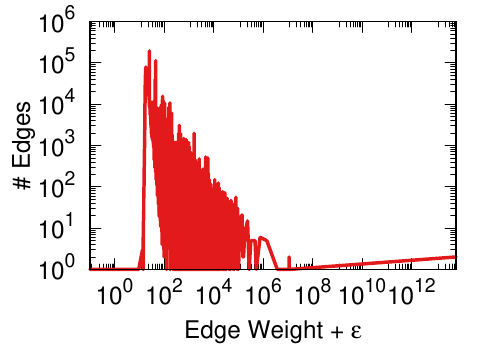}
\includegraphics[width=.475\columnwidth]{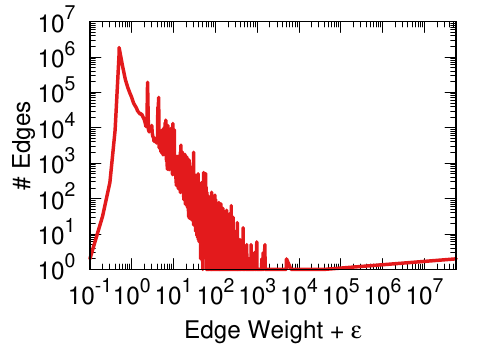}
\includegraphics[width=.475\columnwidth]{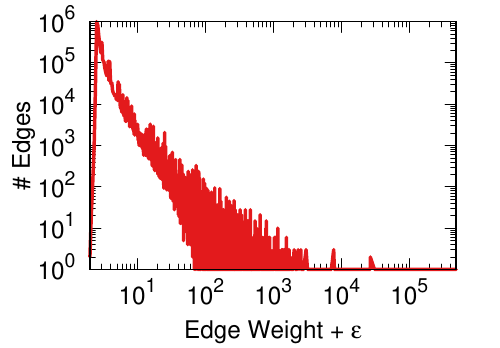}
\includegraphics[width=.475\columnwidth]{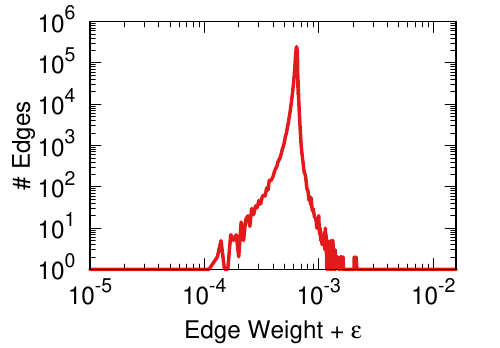}
\caption{The edge weight distributions given different network projection (from left to right columns: simple, hyperbolic, ProbS, YCN) and network backboning (from top to bottom row: naive thresholding, disparity filter, noise corrected) strategies.}
\label{fig:weight-distr}
\end{figure*}

The number of edges is lower than the number of tweets. A tweet generates a user-domain edge only if the user links to the domain in their tweet. As expected, both the user and the domain degree distributions are fat tailed. Figure \ref{fig:degdistr}(left) shows that the most popular domain has been shared by $\sim 3,000$ users, while the average domain only has 9 sharers. On the other hand, the degree distribution is less skewed for users: the most prolific one shared 1,284 domains, while the average one 45.

The projection and backboning task is complicated by the fact that the user-domain bipartite network is degree disassortative. There is a significant negative correlation between the (logged) user and domain degrees of connected nodes: -0.33, whether calculated as a Pearson or a Spearman correlation. Figure \ref{fig:degdistr}(right) shows the joint degree distribution.

\subsection{Threshold Impact}
For this and the following sections, we explore the impact of different projection and backboning techniques. We focus on four projection techniques: simple, hyperbolic, ProbS, and YCN. We use three backboning techniques: naive, disparity filter (DF), and noise-corrected (NC). This means that we have a combination of twelve approaches: all possible combinations of two techniques.

Each projection of the user-domain bitartite network will generate different edge weights in the domain-domain unipartite network, as we discussed in Section \ref{sec:algorithms-projection}. Each backboning technique will re-weight the edges such that one can choose a meaningful threshold to filter edges as we discussed in Section \ref{sec:algorithms-backboning}. Thus, each projection-backboning combination generates a different weight edge distribution. Figure \ref{fig:weight-distr} shows the twelve weight distributions we generated.

The backboning re-weighting influences the distribution the most. Naive thresholding (Figure \ref{fig:weight-distr} top row), which does not re-weight at all, leaves the edge weights broadly distributed. This makes the threshold choice hard for the reasons explained in Section \ref{sec:algorithms-backboning}.

DF backboning (Figure \ref{fig:weight-distr} middle row) spreads all weights in the 0-1 space uniformly, with the exception of a peak around weight 0.63. That is the maximum possible DF weight of an edge whose original weight was one. Since we have a broad weight distribution (see Figure \ref{fig:weight-distr} top row leftmost column), such value is the most popular. This also constrains the threshold choice: either one has to filter out all edges with original weight equal to one, or they have to set the threshold to 0.63, including way too many edges.

NC backboning (Figure \ref{fig:weight-distr} bottom row) spreads all weights on a pseudo-lognormal distribution.

\begin{figure*}
\centering
\includegraphics[width=.6\columnwidth]{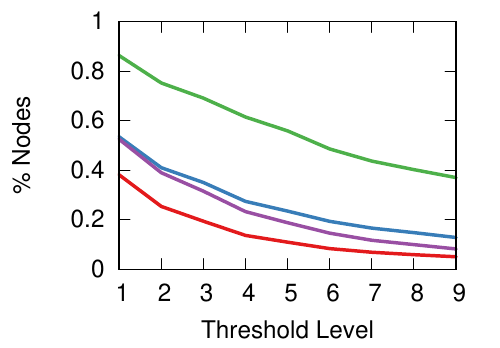}
\includegraphics[width=.6\columnwidth]{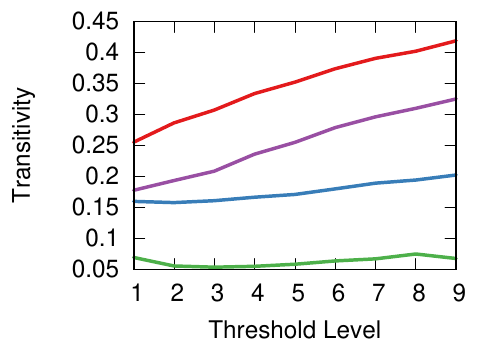}
\includegraphics[width=.6\columnwidth]{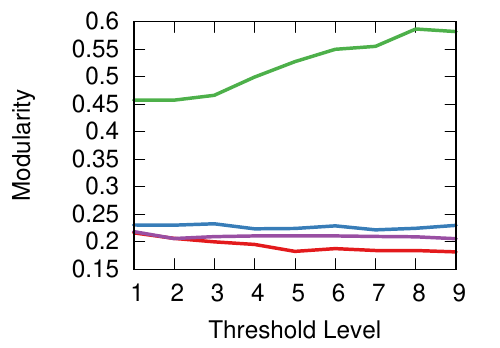}

\includegraphics[width=.6\columnwidth]{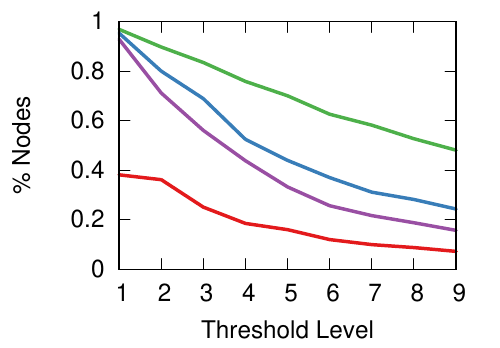}
\includegraphics[width=.6\columnwidth]{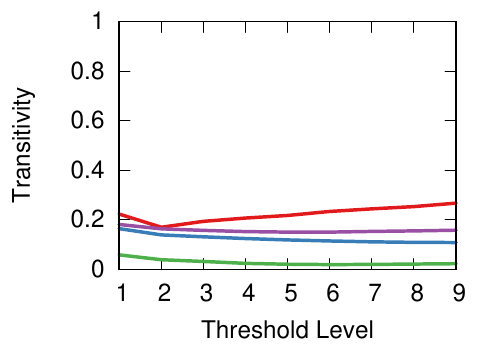}
\includegraphics[width=.6\columnwidth]{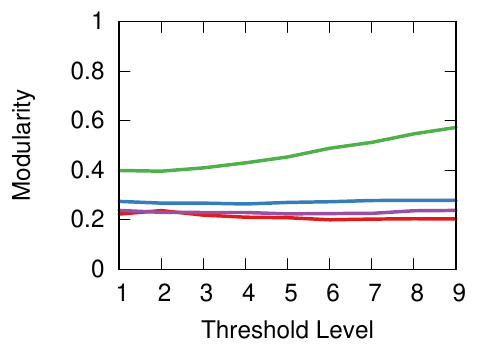}

\includegraphics[width=.6\columnwidth]{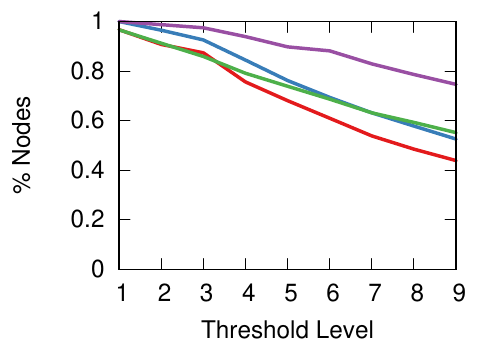}
\includegraphics[width=.6\columnwidth]{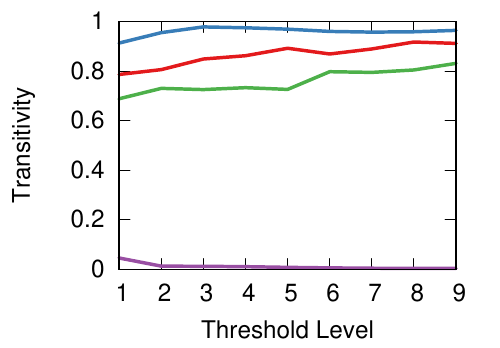}
\includegraphics[width=.6\columnwidth]{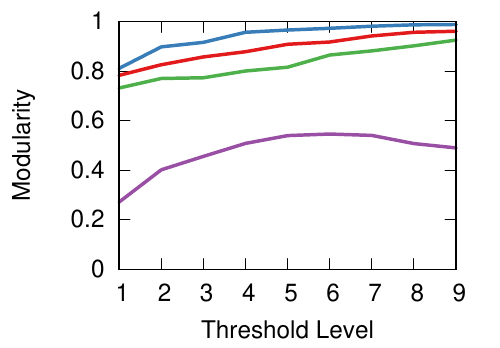}

\includegraphics[width=\columnwidth]{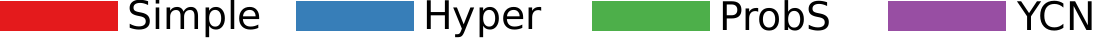}
\caption{The topological characteristics (from left to right: share of nodes with at least degree equal to one, clustering coefficient, and modularity) of networks extracted with different network projection (colors: red = simple, blue = hyperbolic, green = ProbS, purple = YCN) and network backboning (from top to bottom row: naive thresholding, disparity filter, noise corrected) strategies.}
\label{fig:topologies}
\end{figure*}

We now want to study the impact of different threshold levels to the topology of the network. We define nine threshold levels, so that each strategy will return a comparable number of edges. Figure \ref{fig:topologies} shows the result.

The leftmost column of Figure \ref{fig:topologies} shows how many nodes we can keep into the backbone, meaning the number of nodes with degree higher than zero. Ideally, we want to sparsify the network, thus removing as many edges as possible, but we do not want to isolate nodes. This is most easily achieved by NC backboning for all projection techniques and for all threshold levels. 

\begin{figure*}
\centering
\includegraphics[width=.66\columnwidth]{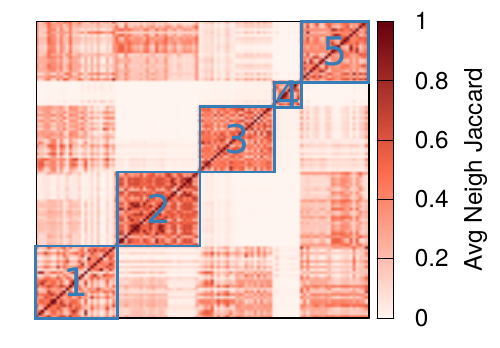}
\includegraphics[width=.66\columnwidth]{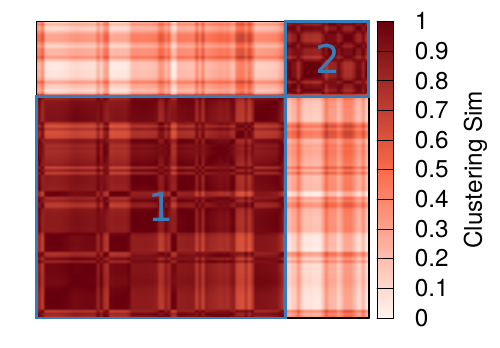}
\includegraphics[width=.66\columnwidth]{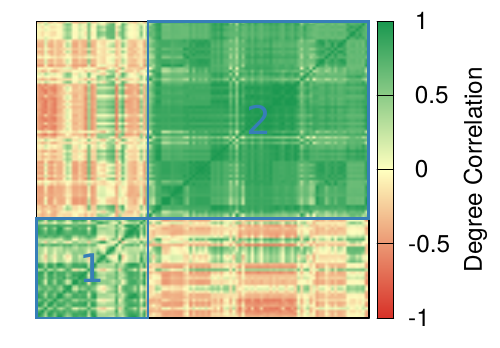}
\caption{The similarities between the networks extracted with different network projection and backbone techniques at different threshold levels (rows and columns).}
\label{fig:similarities}
\end{figure*}

The middle column of Figure \ref{fig:topologies} shows the transitivity of the backbones, namely the share of closed triangles. The results show a high impact of the strategy combination rather than threshold level. The lines are flat, meaning that the average standard deviation in the same strategy combination is low ($\sim 0.028$). However, the average difference across strategies is high, for instant hyperbolic+NC generate almost fully clustered networks, while YCN+NC's transitivity is almost zero.

The leftmost column of Figure \ref{fig:topologies} shows the modularity of communities extracted for each backbone -- i.e how well-separated their communities are \cite{newman2006modularity}. The same discussion holds as in the case of transitivity: low average standard deviation within strategy at different threshold levels ($\sim 0.029$) but wide differences across strategies.

We can conclude that the threshold choice has a limited impact on the topology of the network. However, there are wide differences when choosing different projection and backboning techinques. We now turn our attention to them, with an attempt to classify them.

\subsection{Strategy Impact}\label{sec:impact}
While in the previous section we have shown how the varying the threshold has a limited impact on the resulting network, the simple combination between projection and backboning leaves us with 12 possible strategies to evaluate and, possibly, classify.

Figure \ref{fig:similarities} shows the similarity between average neighbors Jaccard, clustering coefficient, and degree correlation of the networks extracted with different combination of strategies and thresholds (rows and columns). Since we have 12 strategies and 9 threshold levels, we have a space of 108 different approaches. Notwithstanding this large space of possibilities, we obtain a relatively small number of clusters producing similar results. 

Table \ref{tab:clusters} shows how the various strategies are combined in each clusters. It is interesting to observe that, while the average neighborhood Jaccard shows a very complex picture with strategies clustering without any evident pattern, things seems to be clearer when we consider the clusters produced comparing the similarity of clustering coefficient and degree correlation. In those cases the methods groups in two clusters, one smaller containing all the strategies that include NC backboning -- except for the combination with YCN projection -- and a larger one that contains all the other methods. In the case of the similarity based on Degree Correlation the small cluster also contains the strategy ProbS projection + Naive backboning at all threshold levels.

This suggests that, while the neighbourhood structure of the networks resulting from the different strategies can be substantially different, other topological characteristics are less affected by the chosen strategy with the exception of those using Noise Corrected backboning in combination of projecting methods that are not YCN.

\begin{figure*}
\centering
\includegraphics[width=.5\columnwidth]{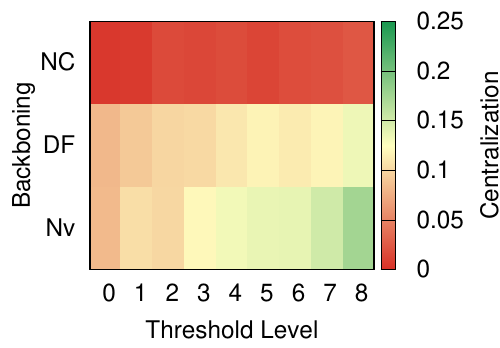}
\includegraphics[width=.5\columnwidth]{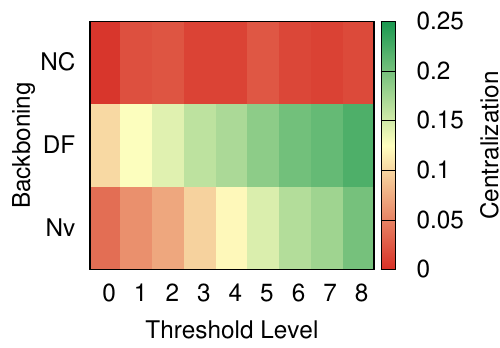}
\includegraphics[width=.5\columnwidth]{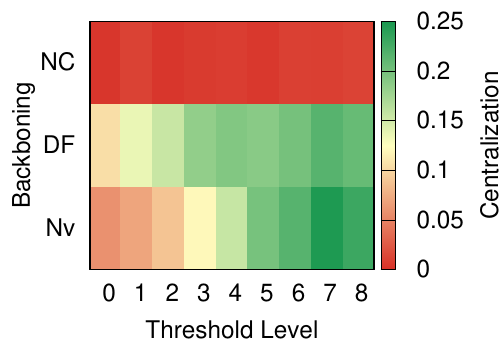}
\includegraphics[width=.5\columnwidth]{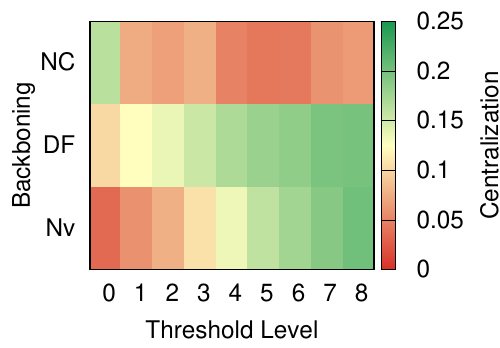}
\caption{The levels of centralization (color: green high, red low) for each projection (left to right: simple, hyperbolic, ProbS, YCN) and backboning (rows) strategy at different threshold levels (columns).}
\label{fig:centralization}
\end{figure*}

This diversity is visible also in Figure \ref{fig:centralization}, where we visualize the level of centralization -- based on betweenneess centrality -- of the networks produced by the different strategies at different levels of thresholding. With all the projection methods, regardless of the threshold level, the networks produced through Noise Corrected backboning are characterized by a consistently low level of centralization -- with the exclusion of YCN. The other backboning methods show a more diverse behaviour largely dependent on the threshold level. For example both a naive approach and disparity filter produce networks with a level of centralization that largely increases following the increment on the level of thresholding. This means that, the fewer edges we include in the network, the more it tends to be dominated by a central hub. 

It is interesting to observe how in the case of YCN projection, while naive backboning and disparity filter produce network that are less centralized with low thresholds, noise corrected shows the opposite behaviour producing a higher level of centralization for low thresholds and a lower level for high threshold values. This apparently odd behaviour should become clear if we think of the specific combination between the assumptions that YCN projection and NC backboning make and that will be discussed in details in the following section.

\begin{table}
    \centering
    \setlength\extrarowheight{2pt}
    \begin{tabular}{|p{15mm}|P{13mm}|p{45mm}|}
    \hline
    Metric & Community & Strategies \\
    \hline
    \multirow{9}{*}{\makecell{Avg. Neigh.\\ Jaccard}} & 1 & \multicolumn{1}{m{45mm}|}{Hyperbolic+Naive (High thresholds); Hyperbolic+DF (All thresholds); Simple/YCN + DF/Naive (Low thresholds).}\\[2pt]
    \cline{2-3}
    & 2 & \multicolumn{1}{m{45mm}|}{Simple/Hyperbolic/ProbS + NC (All thresholds)}\\[2pt]
    \cline{2-3}
    & 3 & \multicolumn{1}{m{45mm}|}{Simple/YCN + Naive/DF (High thresholds)}\\[2pt]
    \cline{2-3}
    & 4 & \multicolumn{1}{m{45mm}|}{YCN + NC (All thresholds)} \\[2pt]
    \cline{2-3}
    & 5 & \multicolumn{1}{m{45mm}|}{ProbS+DF (All thresholds); ProbS+Naive (All thresholds); Hyperbolic+Naive (High thresholds).}\\  \hline 
    \multirow{3}{*}{Clust. Coeff.}  & 1 &  \multicolumn{1}{m{45mm}|}{All remaining strategies (All thresholds)} \\[2pt]
    \cline{2-3}
    & 2 & \multicolumn{1}{m{45mm}|}{Simple/Hyperbolic/ProbS + NC (All thresholds)}\\ \hline 
    \multirow{3}{*}{Degree Corr.}  & 1 & \multicolumn{1}{m{45mm}|}{Simple/Hyperbolic/ProbS + NC (All thresholds); ProbS+Naive(All thresholds)}  \\[2pt]
    \cline{2-3}
    & 2 & \multicolumn{1}{m{45mm}|}{All remaining strategies (All thresholds)} \\
    \hline
    \end{tabular}
    \caption{Projection and backboning strategies clustered according to the similarity of the resulting network structures}
    \label{tab:clusters}
\end{table}

\section{Conclusion}
In this paper we analyze the effects on the topological properties of a network of the combination between different bipartite projections followed by backboning techniques. We build a bipartite network on which we apply four projection and three backboning methods. The analysis points to two results and one lesson that should be kept in mind when we approach bipartite network data with the plan of performing projection and backboning as a preliminary activity.

First: projection and backboning techniques are based on assumptions on the data, on their meaning, and on their properties. Some assumptions might or might not hold in the data we study, e.g. approaches based on resource allocation such as ProbS might be better for some types of bipartite networks. Nevertheless there are cases where the combination between the assumptions made by the projection phase result in largely incompatible data with what is expected from the backboning part. This is the case for YCN projection and noise corrected backboning, where the same normalization approach considering the tendency of connection between originating and receiving nodes would be performed twice, with unexpected results.

Second: despite the large number of possible combinations between projection and backboning, the various strategies cluster in two groups when we take into consideration the topology of the resulting networks. The discriminant factor seems to be the backboning technique. The strategies present in the two clusters produce largely different network topologies when we observe the overall level of centralization. The strategies that involve NC consistently produce decentralized networks, while strategies including disparity filter or naive projections seem to be more sensible to the level of threshold, thus producing more and more centralized networks the fewer edges are included in the projected network.

Projection and backboning always produce a loss in the complexity of the data: this is necessary to improve the explorability of the data and to bring to surface hidden information. Each of the available strategies that we have tested have been shown to perform well under certain assumptions and in specific contexts. But the existence of these two groups, one producing centralized and the second producing decentralized networks, implies that we should be cautious when designing an analysis to study, e.g., centralization. It would be problematic to choose a backboning technique we know induces centralization in the resulting network, if the level of centralization is the objective of our study. A strategy that discourages the creation of centralized networks (such as noise corrected as backboning technique) could be preferable. A transparent use of a selected combination of projection and backboning methods appears to be necessary to avoid problematic methodologically-induced results.

\bibliographystyle{plain}
\bibliography{biblio}

\end{document}